\documentclass[journal]{IEEEtran}
\usepackage[utf8]{inputenc}
\usepackage[T1]{fontenc}
\usepackage{amsmath}
\usepackage{nccmath}
\usepackage{pst-all}	
\usepackage{algorithm}
\usepackage{algorithmic}
\usepackage{setspace}
\usepackage{cite}
\usepackage{graphicx}
\usepackage{footnote}
\usepackage{balance}

\begin{document}

\title{ Global Stock Market Prediction Based on Stock Chart Images Using Deep Q-Network }

\author{ Jinho Lee, Raehyun Kim, Yookyung Koh, and Jaewoo Kang, 
        
\thanks{ All authors are with the Department of Computer Science and Engineering, Korea University, Seoul, Korea (e-mail: jinholee@korea.ac.kr; raehyun@korea.ac.kr; ykko603@korea.ac.kr; kangj@korea.ac.kr) }%
\thanks{Contact: kangj@korea.ac.kr}%
}

\maketitle

\begin{abstract}
We applied Deep Q-Network with a Convolutional Neural Network function approximator, which takes stock chart images as input, for making global stock market predictions. Our model not only yields profit in the stock market of the country where it was trained but generally yields profit in global stock markets. We trained our model only in the US market and tested it in 31 different countries over 12 years. The portfolios constructed based on our model's output generally yield about 0.1 to 1.0 percent return per transaction prior to transaction costs in 31 countries. The results show that there are some patterns on stock chart image, that tend to predict the same future stock price movements across global stock markets. Moreover, the results show that future stock prices can be predicted even if the training and testing procedures are done in different countries. Training procedure could be done in relatively large and liquid markets (e.g., USA) and tested in small markets. This result demonstrates that artificial intelligence based stock price forecasting models can be used in relatively small markets (emerging countries) even though they do not have a sufficient amount of data for training.

\end{abstract}


%
\IEEEpeerreviewmaketitle

\section{Introduction}

\IEEEPARstart{P}{redicting} future stock prices has always been a controversial research topic. In "Efficient Capital Markets\cite{Fama1970},'' Eugene Fama argues that the stock market is highly efficient and the price always fully reflects all available information. He also maintains that technical analysis or fundamental analysis (or any analysis) would not yield any consistent over-average profit to investors. On the other hand, numerous studies from various domains have reported that the stock market is not always efficient and it may be at least partially predictable\cite{Barberis2003,Malkiel2003}. Some classical works in financial economics found and analyzed numerous anomalies inconsistent with EMH \cite{Schwert2003,Abarbanell1997}. Some other works have used technical analysis, which is the study of past stock price and volume, to predict future stock prices and demonstrated its profitability  \cite{Park2007}. In the computer science domain, many studies have analyzed Web data such as Social Networking Service (SNS) messages \cite{Bollen2011}, news articles \cite{Schumaker2009}, or search engine queries \cite{Preis2013}. Some studies found that investors' sentiments from SNS platforms and search query frequency data provide useful information for predicting future stock prices.

One of the other approaches to predicting future stock prices in the computer science field is to build artificial intelligence based models which use machine learning techniques such as Neural Network (NN) \cite{Lecun2015} or Reinforcement Learning (RL) \cite{Sutton1998}. NN and RL are currently among the most commonly used machine learning methods. NN are used for detecting non-linear patterns in raw input data. Many state-of-the-art methods in various domains such as natural language processing, image classification, and speech recognition are based upon Convolutional Neural Network (CNN) or Recurrent Neural Network (RNN) models. The goal of RL is to train an agent to choose the optimal action given the current state. But unlike supervised learning where exact answers are given to a model, an RL agent is trained to maximize cumulative reward in the training process. Many studies have implemented NN based architectures\cite{Atsalakis2009,Takeuchi2013,Krauss2017,Fischer2018} or RL \cite{Deng2016} techniques. These studies demonstrate that among various input variables, such artificial intelligence based models efficiently capture complex non-linear patterns associated with future returns. But most of the previous works mainly focused on building a high performance model optimized on a limited number of securities or composite indexes only in a single country using various input variables such as price, volume, and technical and other financial indicators. None of these works applied their model to the global stock market.

In this work however, we mainly focus on learning patterns that generally yield profit not just in a single country but also in global stock markets using stock chart images which show past daily closing prices and volume as input. For example, let us assume that our model learned some unique patterns from the training data of a single country, and the patterns indicate that the stock price will sharply go up. Then, we need to show that these unique patterns consistently indicate the same future stock price movement (that the stock price will rise) not only in the stock market of the country in which our model was trained but also in many others. Interestingly, as our experimental results show, the investment activities of people from different countries and cultures tend to be similar for certain price/volume patterns, which makes the patterns more robust and reliable. Moreover, existence of these patterns can be used to help train some artificial intelligence based stock price prediction models on a sufficient amount of data from well developed markets (e.g., USA) and apply them in relatively small countries which do not have enough data to train such complex models. To the best of our knowledge, no previous study has been conducted to address this problem. 

We adopt the framework of Deep Q-Network (DQN) \cite{Mnih2015}, which solves the instability problem which is caused by using highly non linear function approximators with Q-learning \cite{Watkins1992}. It uses the following two methods to stabilize the training process: experience replay and parameter freezing. We use same methods in our training process. Our model takes chart images of an individual company as input and chooses one action among Long, Neutral, or Short every day. It receives positive or negative reward based on its action and the subsequent price change of a company. Our model is trained to select the action that will yield maximum cumulative reward given chart images. Using the Q-learning  (which is one class of RL) to train our model has some advantages over using supervised learning. First, like all other RL, our model is trained using reward. Since we are dealing with the stock price prediction problem, assigning  binary labels (e.g., True or False) to actions is insufficient. For example, if a model decides to take a long action, it is desirable to receive 10.0 reward for a +10\% subsequent price change and 1.5 reward for +1.5\%. Only receiving True for both cases does not give any distinction between the two cases. Second, RL uses cumulative reward, not just immediate reward, to train an agent. In most stock price prediction problems, supervised learning models are trained to predict the price (or price change) of the next time step based on the information of the current time step. In supervised learning, one can use different units of time such as months or days to make predictions on longer or shorter terms, but it is quite difficult to consider the time steps following the next time step. But RL can efficiently handle this problem by maximizing cumulative reward using information from not only the next time step but from all subsequent time steps. Finally, in Q-learning, a trained model can make use of an action value, which is the expected cumulative reward of a corresponding action. So when training is done, our model not only knows which action to take but also can predict the amount of profit the action will yield, which enables us to distinguish strong patterns from weak ones.

We conducted numerous experiments on global stock markets. For this work, only five years (Jan.2001-Dec.2005) of US individual stock data are used for training and our model is tested in 31 countries over 12 years (Jan.2006-Dec.2017) after the training period. The results show that our model generally yields more than the average market return in most of the countries. In other words, our model can detect patterns in stock charts that not only yield profit in the country where our model is trained but also in most of other counties as well. Moreover, the results demonstrate that unlike most of the previous works, an artificial intelligence based stock price prediction model does not need to be trained and tested in the same market or country. For example, it is possible to use US market data for training an NN model for Spain or Taiwan. This may help an artificial intelligence based stock price prediction models to be more widely used in emerging markets, some of which are inefficient and have an insufficient amount of data for training models. As the results show, even though our model is trained in the US, it generally yields a considerable amount of profit in other countries. To the best of our knowledge, our artificial intelligence based model, which is trained in only one country, is the first to obtain numerous testing results on global stock markets.

\begin{figure*}[ht]
\centering
\includegraphics[width=1.0\textwidth]{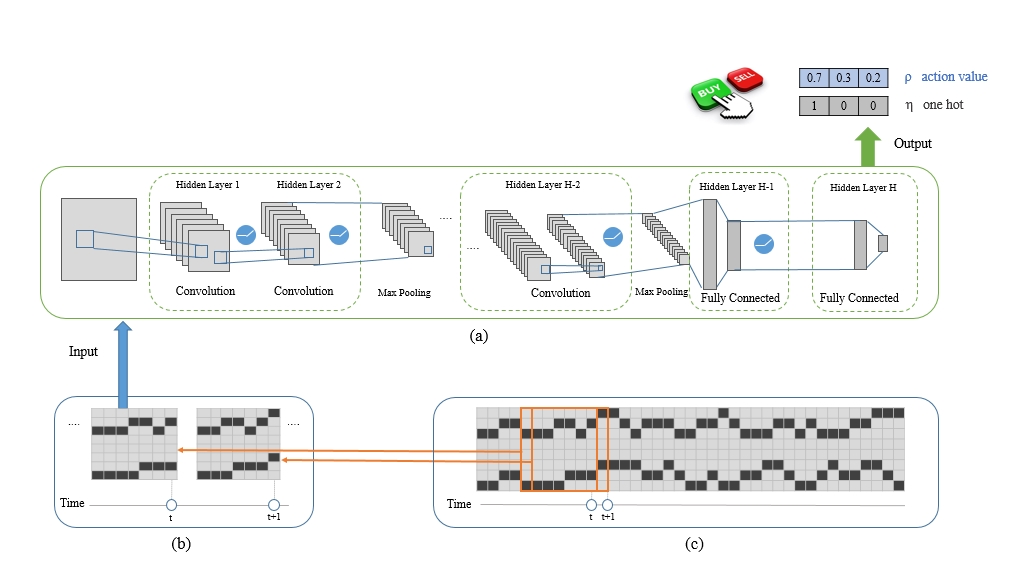}
\caption{ Overview of how our CNN reads an input chart of a single company at a specific time point (time \emph{t}) and outputs the two vectors $\rho$ and $\eta$. (a) Architecture of CNN. It consists of H hidden layers. The last two hidden layers are fully connected layers. (b) Example of a W by W chart image at time \emph{t} and a W by W chart image at \emph{t+1}. For example, if W equals 8 as shown in this figure, our CNN reads as input an 8 by 8 matrix with all elements filled with 0 or 1. Elements filled with the color black corresponds to 1; otherwise, they are all 0. The top part of the matrix represents the relative value of the closing price and the lower half represents volume. Two rows in the middle of the chart are empty (has zero value) to help our CNN to distinguish price from volume. (c) Sequential chart of 39 consecutive days. In this figure, all price volume data are normalized over 39 days for visualization. In other words, for price data, the highest price in 39 days is listed in the first row, and the lowest price is provided in the third row; however, this is only for visualization purposes. In our actual experiments, input data are normalized over W days (horizontal size of chart), and is not normalized over the entire experimental period. }
\label{fig:Fig1}
\end{figure*}

\section{Background}

\subsection{Convolutional Neural Network}

Deep learning or NN is currently one of the most widely used machine learning methods specialized to classify highly non-linear patterns. CNN is one of the NN architectures that was successfully applied to image classification problems. Many state-of-the-art image classification models are based upon CNN architecture. Such models usually takes 2D images as input with three color channels. This input goes through multiple hidden layers. Typically, each hidden layers consists of convolutional layers followed by non-linearity and pooling layers. But in last one or two hidden layers, usually fully connected layers are used with soft-max function. The final output is usually a one-hot vector that corresponds to the label of the input image. Note that in our work, we use CNN as a function approximator in the Q-learning algorithm.

\subsection{Q-learning} Q-learning is one of the most common RL algorithms. Basically, the goal of all RL algorithms is to enable an agent to learn optimal policies, or in other words, train an agent so that it is capable of choosing the action that would give maximum cumulative reward in a given state. In Q-learning, an agent does not directly learn optimal policies; instead, an agent is trained to acquire the optimal action value which is the expected cumulative reward of each action given the current state. So when training is done, the optimal policy of an agent is simply a greedy policy where an agent chooses the action with the maximum action value given the state. To obtain the optimal action value, an agent should iteratively update the action value using the Bellman Equation. An agent chooses action given the current state following behavior policy, and observes reward and next state. Usually in Q-learning, the $\epsilon$-greedy policy is used as a behavior policy, where an agent either chooses a random action with probability $\epsilon$ or acts greedily.

\subsection{Deep Q-Network}
When state representation is simple, the original Q-learning algorithm is proven to converge at optimal behavior. But like in Deep Q-Network (DQN), if the current state is very complex and cannot be represented in table lookup form, one can use the function approximator to efficiently represent the state. The function approximator could be any type of function that maps raw state representations to actions. In our case, CNN is used as the function approximator which maps a state representation (stock chart image) to an action (long, neutral, or short). But naively implementing a non-linear function approximator such as NN is known to be ineffective in real practice because the training process is unstable. DQN addresses this problem using the following two methods: experience replay and parameter freezing. Experience replay is a way to reduce correlations in the sequence of data by storing the latest M experience (input data) in the memory buffer and sampling random batches from the memory buffer at every iteration to take the gradient step. The parameter freezing method temporarily freezes the target parameters during training. To reduce correlations with the target, two sets of parameters are maintained and the target network parameters are updated periodically.

\begin{table*}[ht]

\caption{Data statistics from 31 countries. The first row lists the training set that contains data collected over a 5-year period (2001-2006) from the US, and all other rows list the test set collected over a 12-year period (2006-2018). }
\label{tab:Table1} 
\centering
\begin{tabular}{|l||l||l|l|l|l|l|l|l|}\hline

\hline \textbf{Symbol} & \textbf{Country} & \textbf{Period}  & \textbf{TotCom\#} &  \textbf{N} &  \textbf{Data\#} &  \textbf{Return Avg} & \textbf{Return Std} & \textbf{Excess Rate}  \\

\hline US &  United States  &  2001-2006 & 2792  &  1534 &  1876082 &  0 &  2.931337 &  0.002093\\
\hline US &  United States &  2006-2018 & 2792 &  2061 &  6019248 &  0.05369 &  2.707849 &  0.00162 \\
\hline AUS &  Australia  &  2006-2018 & 960&  485 &  1421920 &  0.040355 &  3.914402 &  0.009042\\
\hline CAN &  Canada &  2006-2018 & 2059&  500 &  1456500 &  0.013443 &  3.426996 &  0.006177\\
\hline CHI &  China &  2006-2018  & 854 &  500 &  1410000 &  0.097583 &  3.084747 &  0.000362\\
\hline FRA &  France &  2006-2018 & 3605 &  500 &  1484000 &  0.020929 &  2.462878 &  0.00128\\
\hline GER &  Germany &  2006-2018 & 4639&  500 &  1475000 &  0.033294 &  2.449741 &  0.001145\\
\hline HK &  Hong Kong  &  2006-2018 & 1674&  500 &  1433000 &  0.025685 &  3.262297 &  0.003275\\
\hline IND &  India &  2006-2018 & 4595 &  500 &  1431500 &  0.052143 &  2.890863 &  0.000822\\
\hline KOR &  South Korea &  2006-2018 & 1493 &  500 &  1440000 &  0.043791 &  3.258717 &  0.000562\\
\hline SWI &  Switzerland &  2006-2018 & 463&  169 &  495736 &  0.026956 &  2.318349 &  0.003326\\
\hline TAI &  Taiwan &  2006-2018 & 1484 & 475 &  1360534 &  0.029255 &  2.431571 &  0.000431\\
\hline UK &  United Kingdom &  2006-2018&  1243 &  500 &  1466500 &  0.021043 &  2.78245 &  0.002242\\
\hline BRA &  Brazil &  2006-2018 & 392 &  94 &  270327 &  0.034498 &  3.109427 &  0.004417\\
\hline DEN &  Denmark &  2006-2018 & 99&  85 &  246942 &  0.026021 &  2.654122 &  0.002369\\
\hline FIN &  Finland &  2006-2018 & 103&  75 & 218495 &  0.020425 &  2.863806 &  0.003391\\
\hline GRE &  Greece &  2006-2018 & 73&   62 & 181693 &  0.018783 &  3.806153 &  0.007584\\
\hline MAL &  Malaysia &  2006-2018 & 764&  100 & 288400 &  0.026428 &  2.881122 &  0.005173\\
\hline NET &  Holland &  2006-2018 & 98&  69 &  206174 &  0.024766 &  2.758522 &  0.005253\\
\hline NOR &  Norway &  2006-2018 & 130&  82 &  239590 &  0.013903 &  3.336327 &  0.003798\\
\hline SIG &  Singapore &  2006-2018 & 315&  99 & 289844 &  0.030436 &  2.678106 &  0.004647\\
\hline SPA &  Spain &  2006-2018 & 136&  87 &  259116 &  0.00173 &  2.612375 &  0.001598\\
\hline SWD &  Sweden &  2006-2018 &488 &  100 &  292600 &  0.028966 &  2.437671 &  0.001411\\
\hline TUR &  Turkey &  2006-2018 & 384&  100 &  299700 &  0.050585 &  2.698828 &  0.000867\\
\hline AUR &  Austria &  2006-2018 & 38&  30 & 85717 &  0.032804 &  2.307613 &  0.001564\\
\hline BEL &  Belgium &  2006-2018 & 96&  80 &  238650 &  0.021966 &  2.297679 &  0.002237\\
\hline IDO &  Indonesia &  2006-2018 & 274&  100 &  285400 &  0.058462 &  3.143907 &  0.00404\\
\hline IRL &  Ireland &  2006-2018 & 27&  18 &  54058 &  0.038105 &  3.008453 &  0.002904\\
\hline ISR &  Israel &  2006-2018 & 224&  100 &  285100 &  0.015192 &  2.599748 &  0.002785\\
\hline ITL &  Italy &  2006-2018 & 284&  100 &  294800 &  0.005149 &  2.494763 &  0.001197\\
\hline POR &  Qatar &  2006-2018 & 30&  25 &  74364 &  0.003484 &  2.84083 &  0.001493\\
\hline TAL &  Thailand &  2006-2018 & 399&  100 &  285000 &  0.049351 &  2.607546 &  0.002526\\

\hline
\end{tabular}
\end{table*}

\section{Method}

\subsection{Overview}

In this subsection, we provide a brief overview of our function approximator CNN. Fig. \ref{fig:Fig1} illustrates how our CNN reads input and outputs action values for an individual company. The term action value refers to the expected cumulative reward of an action. In the early training stage, this action value is meaningless (random) but when training is done properly, it indicates how much profit the corresponding action will yield. As shown in Fig. \ref{fig:Fig1}, our CNN takes a W by W chart image as input at each time step \emph{t}, which shows the daily closing price and volume data of a single company over the last W days. At time \emph{t}, our CNN outputs two length 3 vectors:$\rho$ and $\eta$. Based on these vectors, the action [Long, Neutral, Short] to take at time \emph{t} is decided. Likewise, at time \emph{t+1} (or the next day), our CNN in Fig. \ref{fig:Fig1} reads a stock chart image at time \emph{t+1} and decides which action to take at time \emph{t+1}. The action value vector $\rho$ represents an action value which is the expected cumulative reward of an action [Long, Neutral, Short]. One hot vector $\eta$ is marked as 1 in the same index where $\rho$ has the maximum action value; otherwise, it is marked as 0. Each element of the vectors represents long, neutral, or short action respectively. Thus, for example, the value of \textbf{ $\rho$[3] } at time \emph{t} denotes the expected cumulative reward if our CNN takes the short action at time \emph{t}. For simplicity, we standardized the index of all vectors in this paper to start from one. To sum up, the way in which our CNN operates is simple. It reads a chart at time \emph{t} and chooses the action which has the maximum action value. At time \emph{t+1}, it receives reward based on the action at time \emph{t} and the price change from time \emph{t} to \emph{t+1}. It takes action at time \emph{t+1} in the same way it does at time \emph{t}.

\subsection{Network Architecture}
Fig. \ref{fig:Fig1} (a) shows the overall structure of our CNN used as function approximator.
Our CNN takes a W by W chart image as input and outputs two length 3 vectors: action value vector $\rho$ and one-hot vector $\eta$. The three elements of the vectors represent long, neutral, and short actions, respectively. Each element of $\rho$ corresponds to the expected action value of the action given the current input. Only one element in the one-hot vector $\eta$ is marked as 1 where the $\rho$ has the biggest action value; otherwise, it is marked as 0. The exact architecture of our CNN is as follows. Our CNN takes 32 $\times$32 $\times$1 as input. The input has only 1 channel because it does not need to be colored. Our CNN has six hidden layers. Thereby, H equals 6 in Fig. \ref{fig:Fig1} (a). The first four hidden layers are convolutional layers followed by a Rectifier non-Linearity Unit (ReLU) and the last two hidden layers are fully connected layers. In the fully connected layers, ReLU is implemented only after the fifth layer. Each of the first four hidden layers consists of 16 filters of size 5$\times$5$\times$1, 16 filters of size 5$\times$5$\times$16, 32 filters of size 5$\times$5$\times$16, and 32 filters of size 5$\times$5$\times$32, respectively, all with stride 1, zero padding and followed by ReLU. Right after the second and fourth hidden layers, a max-pooling layer with a 2$\times$2 filter and stride 2 is applied. The last two hidden layers are fully connected layers with 2048$\times$32 and 32$\times$3 parameters, respectively, followed by ReLU except for the final layer. The batch normalization \cite{loffe2015} layer is added in every layer right before ReLU. The parameters are initialized using Xavier initialization\cite{Glorot2010}. The softmax function is not implemented since the output of our CNN is an action value, and not a probability distribution between 0 and 1.

\subsection{Data Description}
We collected daily closing price and volume data from Yahoo Finance. But Yahoo Finance does not provide the list of companies that can be download from the web site, we obtained the list of companies of roughly 40 countries including most of the developed markets from http://investexcel.net/all-yahoo-finance-stock-tickers/. Only for US, we used the list of companies of Russell 3000 index (The first half of 2018). We downloaded the adjusted closing price data to reflect events such as stock splits. Countries that did not have enough valid data were excluded. The data of 30 countries collected over 12 years and data of one country (US) collected over 17 years were downloaded. In each country, we also eliminated companies with noisy data. First, we eliminated companies that had no price data. Second, we also eliminated companies that had an excessive number of days with zero volume (we eliminated the companies that had zero volume for more than 25\% of the entire testing period). Strictly speaking, many days of zero volume may not be considered noise because a company may not trade on some days or in some cases, stocks may be suspended for trading for a certain period of time. But stocks that have been suspended for more than 25\% of the entire testing period are definitely abnormal, and may indicate that the data of the given company are erroneous. Thus, excluding such companies does not undermine the validity of our work.

After downloading and eliminating noisy data, the entire dataset is divided into the training set and test set. The training set contains only US market data collected over a five-year period (Jan.2001-Dec.2005) from approximately 1500 companies that are included in the Russell 3000 Index (The first half of 2018). Only about half of the companies listed in Russel 3000 index in the first half of 2018 had data from Jan.2001 to Dec.2005. 80\% of the training set is actually used to train our model and 20\% is used to optimize the hyperparameters. The test set contains data from 31 countries including the US, which was collected over a 12-year period (Jan.2006-Dec.2017). The test set is further divided into four-year intervals as follows:(2006-2010), (2010-2014), (2014-2018). Every four years, the top N liquid companies are selected from each country for the experiment. A value of 3000, 500 and 100 are initially assigned to N for US, developed countries, and emerging countries, respectively. The values are selected based on market capitalization and the number of available companies in each country. All the available companies were used if the number of valid companies were less than initial N value. When selecting the top N liquid companies, the data collected over 30 business days prior to the first day of every four years were used. For example, the top N liquid companies used from Jan.2006 to Dec. 2009 were selected based on data from Nov. 15, 2005 to Dec. 31, 2005. Not all companies have all 12 years of data. The companies listed in Jan. 2010 have data starting from Jan. 2010. So companies that were listed in the exchange market for the entire four-year period were used for that testing period. In other words, all N companies used in the testing period (Jan.2006-Dec.2009) were listed before Jan.2006 (strictly speaking, Nov. 15, 2005) and were still listed in the exchange market after Dec.2009.

In our experiments, we use daily closing price and volume data downloaded from Yahoo Finance, and we convert the raw data to input data as follows. A single data input (corresponds to a single day of one company) consists of two parts: input chart $S_t^c$ and scalar value $L_t^c$. The superscript \emph{c} and subscript \emph{t} indicate company \emph{c} and time \emph{t}, respectively. The input chart $S_t^c$ is a W by W matrix in which all elements are either 0 or 1. The W by W matrix consists of the last W days of closing price and volume data of a single company. For example, the input chart of day $\emph{t}$ of company $\emph{c}$ contains price and volume data from day $\emph{t-W+1}$ to $\emph{t}$ of company $\emph{c}$. Fig. \ref{fig:Fig1} (b) shows an example of an 8 by 8 input chart. A single column represents a single day. The upper half (from rows 1 to 3) represents the relative value of the closing price for 8 days and the lower half (from rows 6 to 8) represents the relative value of the volume. The two rows in the middle of the chart (rows 4 and 5 in this case) are filled with zeros. The two rows in the middle work like zero padding, and helps our CNN to distinguish price from volume. When price and volume are included in a chart, the values of price and volume are min-max normalized over \emph{W} days. Thus, the highest values of price and volume are listed in the first and sixth rows, respectively, and the lowest values are listed in the third and eighth rows, respectively. $L_t^c$ is a scalar value that represents the price change in percentage from day $\emph{t}$ to $\emph{t+1}$. In other words, $L_t^c$ is simply the daily return of company \emph{c} from day $\emph{t}$ to day $\emph{t+1}$. In Fig. \ref{fig:Fig1}, chart $S_t^c$ is shown as the only input to our CNN because the scalar value $L_t^c$ is used with our CNN output to calculate rewards. But in the actual training/testing process, our CNN receives a W by W matrix of company $\emph{c}$ on day $\emph{t}$ as input $S_t^c$ and outputs an action based on $S_t^c$. The reward for this action is calculated using the scalar value $L_t^c$. Equation (\ref{eq1}) calculates $L_t^c$ where $Prc_t^c$ indicates the closing price of company \emph{c} at time \emph{t}.

\begin{equation} 
\label{eq1}
L_t^c =  100 \times (Prc_{t+1}^c - Prc_t^c) / Prc_t^c
\end{equation} 

\noindent While generating $L_t^c$, we applied two simple methods to help our training. First, we bounded the values between -20\% and 20\% to prevent an excessive rewards from noisy data. Though we tried to remove noise data, there may still be some noisy data. Since (\ref{eq1}) involves division, the value of $L_t^c$ can easily change when an extremely small or potentially incorrect value is assigned to the closing price. Even a small amount of noisy data might affect the entire experiment. By bounding the values of $L_t^c$, we could minimize the impact of such undesirable cases. In addition, we conducted more experiments with less tight bounds (50\%, 100\%) but there was no notable change in the results. Second, in the training set, we neutralized the daily return $L_t^c$ to address the data imbalance problem (in our case, having more positive values than negative values). In other words, the daily return averaged over the entire training set is subtracted from each daily return $L_t^c$. If we sample stock market data for a long period of time, the data usually becomes imbalanced because the market tends to go up. So the data usually has more positive values than negative values. Although the degree of imbalance in the stock market data is not that significant, we found that neutralizing the imbalance improves our training process. The test set is not neutralized. Table \ref{tab:Table1} summarizes the information about the dataset used in our experiments. The column TotCom\# and the column N indicate the number of available companies from each country and N value, respectively. The column Data\# is the total number of data used in our experiments. The column Return Avg is the average daily return (in percentage) of the buy and hold portfolio over the corresponding period. As shown in the first row of Table \ref{tab:Table1}, the return average of the training set is 0 because we neutralized the training set. The column Return Std is the standard deviation of daily returns. The column Excess Rate indicates the percentage of data with the absolute value of $L_t^c$, which originally had a value larger than 20\% before bounding. 

Our training and test sets are in the form of a matrix. For example, the training set consists of N $\times$ T data points where N is $\approx$1500 and T is $\approx$1000 (number of business days in four years which is 80\% of entire training set). The test set is formatted in the same way with different values of N and T.

\subsection{Training Process}

\begin{algorithm}
\caption{ Training algorithm }
\begin{algorithmic}[1]
\STATE Initialize memory buffer to capacity M \\
\STATE Initialize network parameters $\theta$  \\
\STATE Initialize target network parameters $\theta^*$ = $\theta$   \\
\STATE Initialize $\epsilon$ = 1 \\

\FORALL{  \emph{b = 1,maxiter } }
    \STATE set \emph{c} $\leftarrow$ random valid company index  \\
    \STATE set \emph{t} $\leftarrow$ random valid date index  \\

    \STATE with probability $\epsilon$ set $a_{t-1}^c$ $\leftarrow$  random \\
    \STATE otherwise $a_{t-1}^c$ $\leftarrow$  $\underset{a}{\max} Q(s_{t-1}^c,a;\theta)$ 

    \STATE with probability $\epsilon$ set $a_{t}^c$ $\leftarrow$ random \\
    \STATE otherwise $a_{t}^c$ $\leftarrow$  $\underset{a}{\max} Q(s_{t}^c,a;\theta)$ 

    \STATE S  $\leftarrow$ \space $s_t^c$ \\
    \STATE A  $\leftarrow$ $a_t^c$ \\
    \STATE R  $\leftarrow$ $a_t^c$ $\times$ $L_t^c$ - P $\times$ |$a_t^c$ - $a_{t-1}^c$| \\
    \STATE S' $\leftarrow$ $s_{t+1}^c$ \\

    \STATE set $e_b$ $\leftarrow$ \big\{S,A,R,S'\big\} \\
    \STATE store experience $e_b$ in memory buffer \\
    
    \IF{   $\epsilon > \epsilon_m $ }
        \STATE $\epsilon \leftarrow \epsilon \times 0.999999 $ \\
    \ENDIF
    
    \IF{  memory buffer is full }
        \STATE delete one oldest experience from memory buffer \\
    \ENDIF
    
    \IF{b \% B == 0}
        \STATE random sample minibatch of size $\beta$ from memory buffer \\
        \STATE $Loss \leftarrow 0$ \\
        \FORALL{  \emph{k in minibatch } }
        
            \STATE set $S_k, A_k, R_k, S_k'$ $\leftarrow$ from $e_k$ \\
    		\STATE $Loss \leftarrow Loss + [R_k+\gamma\underset{a}{\max}Q(S'_k,a;\theta^*)-Q(S_k,A_k;\theta)]^2$ \\
        \ENDFOR
        \STATE $Loss \leftarrow  Loss / \beta $ \\
        \STATE perform gradient step to minimise Loss with respect to parameters $\theta$ \\
    \ENDIF
    
    \IF{ b \% (B*C) == 0}
        \STATE set $\theta^* \leftarrow \theta$ \\
    \ENDIF
    
\ENDFOR
\end{algorithmic}
\end{algorithm}

\begin{table*}[ht]

\caption{List of hyperparameters mentioned in the paper. Hyperparameter optimization is done using 20\% of the training set.  }
\label{tab:Table2} 
\centering
\begin{tabular}{|l||l||l|l|}\hline
\textbf{Hyperparameter}  & \textbf{Description} & \textbf{Value}   \\
\hline \emph{maxiter} & Maximum number of iterations  & 5,000,000      \\
\hline \emph{learning rate} & The learning rate used by Adam optimizer & 0.00001      \\
\hline \emph{ $\epsilon_m$} & Minimum value of $\epsilon$ & 0.1 \\
\hline \emph{W} & Horizontal and vertical size of input matrix   & 32      \\
\hline \emph{M} & The capacity of memory buffer  & 1,000       \\
\hline \emph{B} & Update interval of parameters $\theta$   & 10      \\
\hline \emph{C} & Update interval of parameters $\theta^*$  & 1,000      \\
\hline \emph{P} & The transaction penalty while training  & 0.05      \\
\hline \emph{$\gamma$} & The discount factor  & 0.99      \\
\hline \emph{$\beta$}  & Batch size  & 32      \\
\hline
\end{tabular}
\end{table*}

The standard Q-learning algorithm is based on the Bellman equation, and iteratively updates its action value based on the assumption that if an action value is optimal then it satisfies the Bellman equation. The Bellman equation defines the relationship between the current action value $Q(s,a)$ and the subsequent action value $Q(s',a')$. The loss function is derived from this equation. Our training process uses the following two methods of DQN: experience replay and parameter freezing. Our loss function is defined in (\ref{eq2}). We use the Adam optimizer \cite{Kingma2014} to perform a gradient step on \emph{ Loss($\theta$) } with respect to parameters $\theta$. For better understanding, the batch size is omitted in (\ref{eq2}) so the loss function can be interpreted as loss calculated from a single experience.

\begin{equation} 
\label{eq2}
Loss(\theta) = [ r +  \gamma \underset{a'}{\max} Q(s',a';\theta^*) - Q(s,a;\theta) ]^2 
\end{equation} 
\noindent where \emph{s}, \emph{a}, \emph{r}, \emph{s'}, and \emph{a'} refer to current state, action, reward, subsequent state, and subsequent action, respectively, and $\gamma$ denotes the discount factor. New symbols are used to maintain consistency with the standard Q-learning algorithm used in previous works. In Fig. \ref{fig:Fig1}, state \emph{s} and action \emph{a} correspond to input chart and output action at \emph{t}, respectively. Likewise, subsequent state \emph{s'} and action \emph{a'} also refer to input chart and output action at time \emph{t+1}, respectively. As mentioned earlier, output $\rho$ in Fig. \ref{fig:Fig1}, which is the output of our CNN, is the action value vector of each element which corresponds to each action [Long, Neutral, Short], respectively. The term $Q(s,a;\theta)$ is a scalar value that represents the action value of action \emph{a} given state \emph{s} using our CNN parameterized by $\theta$. Thus, given state $\emph{s}$, if action \emph{a} is \emph{Short}, then $Q(s,a;\theta)$ exactly corresponds to output $\rho$[3] from our CNN parameterized by $\theta$. Here, the network parameters $\theta$ and target network parameters $\theta^*$ are maintained throughout the training process to implement the parameter freezing method. Both $\theta$ and $\theta^*$ are randomly initialized with the same value in the beginning of the training stage. In the original version of the parameter freezing method, the optimizer performs a gradient step on \emph{Loss($\theta$)} with respect to the network parameters $\theta$ at every iteration, and no gradient step is performed with respect to the target network parameters $\theta^*$. Target network parameters $\theta^*$ are only updated at every \emph{C} iteration by copying parameters $\theta$ to $\theta^*$.

Although our training algorithm is based on the standard Q-learning algorithm, our algorithm differs in the following ways. Unlike the standard Q-learning algorithm, our algorithm needs information about previous actions to calculate the current reward. Reward $r_t^c$ is calculated as below. Superscript \emph{c} and subscript \emph{t} are added in (\ref{eq3}) and denote company \emph{c} and time \emph{t}, respectively.
\begin{equation}
\label{eq3}
r_t^c = a_t^c\times L_t^c - P\times |a_t^c - a_{t-1}^c|
\end{equation}
\noindent where $r_t^c$, $L_t^c$ and $a_t^c$ are reward, next day return, and action of company \emph{c} at time $\emph{t}$, respectively. Scalar value $L_t^c$ in (\ref{eq3}) and that in (\ref{eq1}) are exactly the same term. Also, $\emph{P}$ denotes transaction penalty. Our model assigns a value of 1, 0 or -1 to $a_t^c$ for long, neutral, or short actions respectively, for company \emph{c} at time $\emph{t}$. Thus, we can interpret the first term on the left side of (\ref{eq3}) as the earned profit by choosing action given state. The second term on the right side of (\ref{eq3}) refers to transaction costs when the model changes position at time $\emph{t}$. The transaction costs were not considered ($\emph{P}$ equals zero) in the testing stage but the model is given some penalty when it changes position in the training stage. Without some penalty, the model could change positions too frequently, which would incur high transaction costs in real practice. Equation \ref{eq3} indicates the model needs to know the previous action ( $a_{t-1}^c$ ) to calculate the current reward. The previous action $a_{t-1}^c$ given the previous state is also chosen by implementing the $\epsilon$-greedy policy. Unlike in the standard Q-learning method, in our method, the next state is not affected by the current action. Thus, when performing experience replay, our training algorithm needs to obtain the previous state and implement the $\epsilon$-greedy policy to obtain the previous action. 

Next, we modified the experience replay introduced in the previous work. First, our model not only samples random batches from the memory buffer to take a gradient step on the loss function but it also randomly generates an experience at every iteration to store it in the memory buffer. Second, our model updates parameters $\theta$ every \emph{B} iteration, and not every iteration like the original version. In other words, our model stores an experience in the memory buffer at every iteration, updates the network parameters $\theta$ at every \emph{B} iteration by taking a gradient step on the loss function, and updates the target network parameters $\theta^*$ at every \emph{B $\times$ C} iteration by copying $\theta$ to $\theta^*$. We modified the original version of experience replay to prevent our model from updating parameters $\theta$ for too many iterations with experiences generated from only few companies. As mentioned earlier, we use 80\% of our entire training set to actually train our model; our training set contains data on approximately 1500 companies, which was collected over 1000 days (total $\approx$1,500,000). The original version of experience replay generates experiences and stores them in the memory buffer by the order of input sequence (one company at a time). Assuming that the size of the memory buffer is 1000, the memory buffer has experiences from only one or two companies over the entire training period. It will take approximately 1000 iterations to observe a experience generated from new company. Randomly generating experiences and taking a gradient step at every \emph{B} iteration are done to help our model use many experiences uniformly generated from the entire training set.

The training algorithm generates experience $e_b$ at \emph{b}th iteration and stores it in the memory buffer. Experience is simply a tuple of the current state, action, reward, and subsequent state, i.e., (S, A, R and S'). The algorithm first randomly selects a data index (\emph{c} and \emph{t}) from the training set. Next, the $\epsilon$-greedy policy (either randomly selects an action with probability $\epsilon$ or acts greedily) is used as the behavior policy on state $s_{t-1}^c$ and $s_{t}^c$ to obtain the previous action $a_{t-1}^c$ and the current action $a_{t}^c$. When implementing the behavior policy, value $\epsilon$ is initialized to 1 and gradually decremented until it reaches the minimum value $\epsilon_m$. Rewards are calculated based on previous and current actions and $L_t^c$, then the tuple (S,A,R and S') is assigned to experience $e_b$. The experience $e_b$ is stored in the memory buffer. At every \emph{B} iteration, the minibatch of size $\beta$ is randomly sampled from the memory buffer and used to calculate \emph{Loss}. A gradient step is taken to minimize \emph{Loss} with respect to parameters $\theta$. The target network parameters $\theta^*$ are updated every \emph{B} $\times$ \emph{C} iteration. The full training algorithm is stated in Algorithm 1. Also, the list of hyperparameters mentioned in this paper is provided in Table \ref{tab:Table2}.

\subsection{Source Code Availability}
Our source code used in our experiment is available at https://github.com/lee-jinho/DQN-global-stock-market-prediction/. Since the converted data (chart image data generated from raw data) that was used as input data in our experiments is too large to be uploaded to the online repository, only the sample data is available. The entire dataset can be provided by the corresponding author by request.

\begin{table*}[ht]
\caption{ Result of the market neutral portfolio on a 4-year interval prior to transaction cost. The column Avg TR Number denotes the average number of transactions of one company per year. The last row lists the overall average of 31 countries.}
\label{tab:Table3} 
\centering
\begin{tabular}{|l||l||*{6}{c|}}
\hline
 & & \multicolumn{2}{ c |} {\textbf{Return (\%) 2006 - 2010}} 
&\multicolumn{2}{ c |} {\textbf{Return (\%) 2010 - 2014}} 
&\multicolumn{2}{ c |} {\textbf{Return (\%) 2014 - 2018}} \\
\hline Symbol & Avg TR Number  & per TR  & Annual  & per TR  & Annual  & per TR  & Annual  \\
\hline US   & 89.75  &  0.21 &  20.64 &  0.10  &  9.39 &  0.05  & 4.11 \\
\hline AUS  & 87.67  &  0.34  & 32.38  &  0.19  & 18.13  &  0.12  & 11.35 \\
\hline CAN  & 79.83  &   0.9  & 106.54  &  0.85  & 100.39  &  0.72  & 75.99 \\
\hline CHI  & 84.38  &  0.92  & 120.95  &  0.41  & 43.94  &  0.16  &  15.2 \\
\hline FRA  & 88.04  &  0.07  &  5.97  &     0  &  0.25  & -0.02  & -2.03 \\
\hline GER  & 89.17  &  0.22  & 21.65  &  0.17  & 15.93  &  0.12  & 10.94 \\
\hline HK   & 83.62  &  0.52  & 55.25  &  0.27  & 25.97  &  0.19  & 18.28 \\
\hline IND  & 86.88  &  0.09  &  8.61  &  0.02  &  1.37  &  0.01  &  0.88 \\
\hline KOR  & 84.96  &  0.06  &  5.45  &  0.05  &  4.51  &  0.07  &  6.01 \\
\hline SWI  & 84.0   &  0.51  & 54.75  &   0.4  & 37.93  &  0.35  &  35.4 \\
\hline TAI  & 88.04  &  0.14  & 13.93  &  0.03  &  2.58  &  0.03  &  2.84 \\
\hline UK   & 81.46  &  0.16  & 13.72  &  0.22  & 20.38  &  0.12  & 10.83 \\
\hline BRA  & 87.5   &  0.86  & 110.43  &  0.23  & 22.91  &  0.12  & 11.45 \\
\hline DEN  & 88.12  &  0.52  & 58.33  &  0.59  & 68.27  &  0.26  & 26.16 \\
\hline FIN  & 90.21  &  0.84  & 113.96  &  0.74  & 95.47  &  0.44  & 47.03 \\
\hline GRE  & 87.29  &  0.37  & 39.79  &  0.79  & 103.85  &  0.83  & 95.87 \\
\hline MAL  & 84.92  &  0.71  & 84.43  &  0.45  & 47.15  &  0.42  & 42.64 \\
\hline NET  & 91.12  &   0.5  & 55.95  &  0.58  & 68.36  &  0.32  &    33 \\
\hline NOR  & 89.12  &  0.42  & 45.13  &  0.59  & 69.08  &   0.3  & 30.32 \\
\hline SIG  & 86.33  &  1.05  & 147.61  &  0.32  & 32.59  &  0.29  & 28.52 \\
\hline SPA  & 88.96  &  0.16  & 14.87  &  0.36  & 36.53  &   0.2  &  18.1 \\
\hline SWD  & 91.67  &  0.19  & 18.84  &  0.11  & 10.79  &  0.07  &  6.18 \\
\hline TUR  & 94.12  &  0.09  &  8.53  &  0.06  &  5.87  &  0.05  &  4.62 \\
\hline AUR  & 93.46  &  0.29  & 32.29  &  0.21  & 22.18  &  0.11  & 10.92 \\
\hline BEL  & 92.67  &  0.43  & 49.99  &   0.4  & 47.29  &  0.26  & 27.46 \\
\hline IDO  & 84.83  &  0.58  & 65.25  &  0.24  & 24.74  &  0.14  & 13.12 \\
\hline IRL  & 96.79  &  0.46  & 56.31  &  0.54  & 70.09  &  0.33  & 36.83 \\
\hline ISR  & 87.29  &  0.22  & 21.42  &  0.09  &  9.19  & -0.06  &  -5.8 \\
\hline ITL  & 93.96  &  0.13  &  12.7  &  0.18  & 18.38  &  0.13  & 13.02 \\
\hline POR  & 96.21  &  0.18  & 18.56  &  0.55  & 72.03  &  0.48  & 56.82 \\
\hline TAL  & 89.04  &  0.34  & 36.74  &  0.12  & 11.85  &  0.22  & 23.32 \\
\hline
\hline Average & 88.43&	0.45&	53.99&	0.33&	38.28	&0.22&	23.35  \\
\hline
\end{tabular}
\end{table*}

\section{Experiments}

\subsection{Portfolio Construction}

Our CNN introduced in previous section basically takes a single chart image as input and outputs an action value for a single company at time \emph{t}. But in experiments, we have to deal with more than one company. To deal with more than one company, we construct a length N portfolio vector $\alpha$ which satisfies $ \sum_{c=1}^N |\alpha[c]| = 1.0 $ based on the output vectors of our CNN. N is the total number of companies. At time \emph{t}, our CNN produces the predictions for N companies instead of one. The portfolio is constructed based on the following N outputs: $\rho_c$ and $\eta_c$ where $1\leq c \leq N$. Thus, the portfolio for N companies is reconstructed every day as done for a single company. $\alpha[c]$ represents the portion of the total asset that should be invested into company \emph{c}. Vectors $\rho_c$  and $\eta_c$  also represent an action value and one hot vector for company \emph{c}, respectively. Note that vector $\alpha$ can have a negative value, which means a short position was taken on the company. For example, assuming that the total asset is 1.0, $\alpha[c]$= -0.008 indicates that our CNN is taking a 0.008 short position on company \emph{c} at time \emph{t}. There may be multiple ways to assign weights to each company even if we use the same output of our CNN. In our experiments, we use two methods that are widely used in previous works and real practice.

\begin{table*}[ht]
\caption{Comparison of the average annual return of the top/bottom K portfolios (K=5, 10, 20) with the average annual market return (the column avg) prior to transaction costs. The average annual market return is the return of the buy and hold portfolio with asset uniformly distributed to N companies.  }
\label{tab:Table4} 
\centering
\begin{tabular}{|l||*{12}{c|} }
\hline
&\multicolumn{4}{ c |} {\textbf{Annual Return (\%) 2006 - 2010}} 
&\multicolumn{4}{ c |} {\textbf{Annual Return (\%) 2010 - 2014}} 
&\multicolumn{4}{ c |} {\textbf{Annual Return (\%) 2014 - 2018}} \\
\hline Symbol & K=5 & K=10 & K=20 & avg & K=5 & K=10 & K=20 & avg & K=5 & K=10 & K=20 & avg   \\
\hline US   & 73.27 & 50.6 & 33.83 & 7.22 & 33.95 & 24.44 & 16.41 & 21.8 & 12.91 & 9.66 & 6.31 & 11.25 \\
\hline AUS   & 97.21 & 76.58 & 57.77 & 13.27 & 44.95 & 36.01 & 27.53 & 7.27 & 26.38 & 22.42 & 16.69 & 10.24 \\
\hline CAN  & 530.01 & 365.83 & 228.41 & 5.72 & 531.17 & 349.42 & 214.83 & 2.82 & 343.17 & 239.42 & 153.84 & 1.61 \\
\hline CHI  & 652.78 & 427.44 & 257.74 & 52.79 & 158.47 & 114.31 & 79.19 & 0.44 & 48.86 & 36.33 & 25.28 & 20.35 \\
\hline FRA  & 8.72 & 8.15 & 6.56 & -1.07 & -2.99 & -2.28 & -1.12 & 8 & -10.34 & -6.46 & -3.29 & 8.88 \\
\hline GER  & 67.91 & 52.69 & 38.88 & 2.03 & 53.39 & 38.83 & 26.07 & 12.71 & 36.08 & 26.63 & 18.88 & 10.41 \\
\hline HK   & 226.4 & 155.79 & 101.14 & 17.29 & 82.71 & 61.36 & 43.95 & -0.07 & 58.37 & 44.98 & 32.38 & 2.19 \\
\hline IND   & 30.48 & 19.76 & 14.07 & 19.62 & 5.15 & 4.27 & 2.61 & -4.27 & 1.56 & 1.56 & 2.05 & 24.23 \\
\hline KOR   & 20.87 & 15.25 & 9.58 & 13.16 & 10.15 & 9.27 & 7.67 & 9.63 & 16.59 & 12.38 & 9.62 & 10.28 \\
\hline SWI   & 235.31 & 156.86 & 101.19 & 2.37 & 109.87 & 84.32 & 67.53 & 7.52 & 129.31 & 94.43 & 62.09 & 10.24 \\
\hline TAI  & 37.51 & 32.9 & 25.33 & 21.08 & 2.65 & 4.2 & 3.54 & 1.18 & 10.16 & 6.96 & 3.62 & 1.8 \\
\hline UK  & 45.76 & 33.12 & 24.18 & -0.62 & 61.68 & 47.43 & 35.22 & 12.35 & 29.23 & 25 & 18.22 & 4.21 \\
\hline BRA   & 526.32 & 375.28 & 229.75 & 21.67 & 96.84 & 67.06 & 38.36 & -0.22 & 29.38 & 21.72 & 18.47 & 6.99 \\
\hline DEN   & 334.31 & 189.9 & 111.45 & -3.56 & 274.01 & 197.13 & 131.86 & 8.72 & 87.98 & 62.96 & 45.46 & 12.75 \\
\hline FIN   & 550.63 & 386.36 & 235.23 & 0.54 & 524.55 & 306.07 & 199.99 & 4.99 & 148.24 & 126.2 & 84.45 & 9.2 \\
\hline GRE  & 108.82 & 95.54 & 66.36 & -1.81 & 365.98 & 292.18 & 188.92 & 5.73 & 444.16 & 350.02 & 226.39 & 10.26 \\
\hline MAL  & 427.61 & 303.35 & 166.39 & 9.52 & 192.24 & 133.81 & 89.49 & 11.3 & 158.57 & 109.59 & 77.21 & -0.8 \\
\hline NET  & 186.72 & 143.99 & 93.57 & 1.64 & 241.76 & 169.39 & 123.16 & 4.79 & 77.5 & 71.12 & 51.36 & 11.48 \\
\hline NOR  & 187.83 & 110.95 & 81.31 & 5.62 & 261.36 & 188.23 & 133.27 & 6.12 & 111.11 & 64.78 & 51.85 & 0 \\
\hline SIG  & 1,021.09 & 622.37 & 333.39 & 16.99 & 116.24 & 85.06 & 57.06 & 6.51 & 96.97 & 73.18 & 51.12 & -0.37 \\
\hline SPA  & 40.29 & 25.69 & 22.17 & -3.98 & 125.39 & 97.02 & 64.82 & 0.58 & 45.88 & 32.87 & 26.69 & 3.42 \\
\hline SWD  & 57.75 & 45.08 & 36.44 & 3.52 & 30.29 & 21.48 & 15.96 & 9.21 & 19.5 & 15.79 & 11.25 & 9.2 \\
\hline TUR  & 21.68 & 19.18 & 16.26 & 9.82 & 15.63 & 12.69 & 10.14 & 9.59 & 2.59 & 6.98 & 6.53 & 18.75 \\
\hline AUR  & 101.34 & 80.69 & 52.04 & 4.12 & 44.42 & 38.15 & 31.16 & 10.79 & 56.89 & 30.42 & 19.84 & 8.93 \\
\hline BEL  & 225.81 & 142.14 & 92.87 & -2.59 & 166.36 & 125.16 & 88.44 & 8.69 & 87.67 & 69.87 & 48.87 & 9.44 \\
\hline IDO  & 241.51 & 206.58 & 131.32 & 23.93 & 68.8 & 55.79 & 43.15 & 18.02 & 27.97 & 35.05 & 24.8 & 2.18 \\
\hline IRL  & 178.87 & 178.87 & 92.51 & 3.07 & 397.81 & 397.81 & 165.08 & 19.68 & 125.96 & 102.13 & 54.36 & 5.27 \\
\hline ISR  & 59.85 & 41.45 & 31.96 & 8.3 & 11.67 & 11.17 & 9.09 & 1.94 & -7.28 & -6.94 & -4.34 & 1.15 \\
\hline ITL  & 31.77 & 24.07 & 18.17 & -8.14 & 60.28 & 43.62 & 32.39 & 4 & 40.85 & 29.08 & 27.12 & 8 \\
\hline POR  & 26.46 & 41.66 & 34.75 & -2.07 & 277.57 & 194.36 & 127.78 & 0.95 & 136.47 & 139.39 & 79.77 & 2.92 \\
\hline TAL   & 117.63 & 93.61 & 65.71 & 4.86 & 46.26 & 38.12 & 21.45 & 21.57 & 63.75 & 51.14 & 38.7 & 11.04 \\
\hline
\hline
Average    & 208.79&145.86&90.66&7.88&142.21&104.71&67.58&7.49&79.24&61.25&41.47&7.92\\ \hline

\end{tabular}
\end{table*}

\subsection{Market Neutral Portfolio}

First, we use the market neutral portfolio which takes the same position on each side (long, short) every day. In other words, the portfolio satisfies $ \sum_{c=1}^N \alpha_n[c] = 0 $ every day. The term "neutralize" could be interpreted as making the average zero. The subscript \emph{n} is added to represent the market neutral portfolio. Because the market neutral portfolio theoretically has Beta of zero, which means it has no correlation with the market average return, the market neutral portfolio is theoretically free from market risk, and its performance is unaffected by the market average return. One hot vector $\eta_c$ is used when constructing the market neutral portfolio vector $\alpha_n$. Steps to create the market neutral portfolio are described in Algorithm 2. First, for company \emph{c}, a scalar value of 1, 0, or -1 is assigned to  $\alpha_n[c]$ for long, neutral, and short actions, respectively, based on vector $\eta_c$. Then, the mean of vector ($\mu_n$ = $\frac{1}{N} \sum_{c=1}^N \alpha_n[c]$) is subtracted from each element of the vector $\alpha_n$. Finally, each element of $\alpha_n$ is divided by the sum of the absolute value of the vector's element ($\Sigma_n$ = $\sum_{c=1}^N |\alpha_n[c]| $) to make sure that the portfolio satisfies $ \sum_{c=1}^N |\alpha_n[c]| = 1.0 $ every day. 

Table \ref{tab:Table3} shows per transaction (the column per TR) and annual return (the column Annual) of our market neutral portfolio. Our test results clearly show that our approach generally performs well in most of the stock markets worldwide during most of the testing periods. Market neutral portfolio covers almost all the companies over the entire testing period. In other words, the portfolio assigns asset (at least a small percentage) to almost all the candidate companies. This demonstrates that the profit does not come from a small number of extremely well performing companies, and our model which has the ability to detect profitable patterns generally works well in many different countries.  Except for a few periods, our model generally yields a $\approx$0.1 to 1.0 percent return per transaction or about a $\approx$10 to 100 percent return per annum prior to transaction costs in developed countries as well as in emerging countries. Interestingly, although our CNN is trained in only the US market, it performs much better in other countries, especially in emerging countries, than in the US. The results empirically prove that there are profitable patterns in stock charts and our model is capable of identifying those patterns and these patterns consistently indicate the same future stock price movement not only in certain country but also in many other global markets.

\begin{figure*}[ht]
\centering
\includegraphics[width=0.9\textwidth]{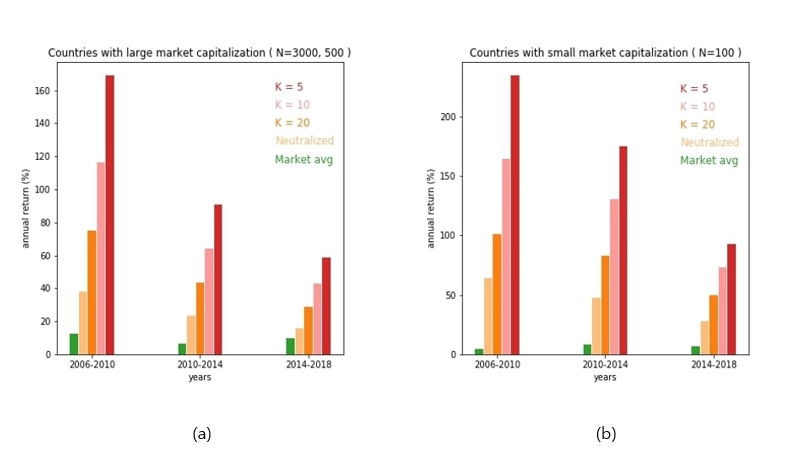}
 \caption{Comparison of the average annual returns of the market neutral portfolio and top/bottom K portfolio with the annual market average return. The graph (a) shows the average results of countries, with relatively large market capitalization, where an initial N value of 3000 or 500 is used in the experiments (12 countries total). The graph (b) shows the average results of other countries (19 countries total), which were obtained by our model which used an initial N value of 100. }
\label{fig:Fig2}
\end{figure*}

\subsection{Top/Bottom K Portfolio}

Next, we use the top/bottom K portfolio, which takes a position only when the signal is strong. In other words, our CNN takes a long position for the top K\% of companies, a short position for the bottom K\% of companies, and no position for the others each day based on vector $\rho_c$, which is another output of our CNN. Note that this portfolio also satisfies $ \sum_{c=1}^N \alpha_s[c] = 0 $ (market neutral). The difference is that the top/bottom K portfolio distributes its asset to only 2 $\times$ K\% companies. To construct this portfolio, first, subtract $\rho_c$[3] from $\rho_c$[1] and use this value to decide which company to take position. Note that each element of vector $\rho_c$ represents the action value of corresponding action. If we take a closer look, the value ($\rho_c$[1] - $\rho_c$[3]) is the difference between the expected cumulative return of the long action and the expected cumulative return of the short action of company \emph{c} at time \emph{t}. Intuitively, this value indicates how much the stock price of company \emph{c} will increase at time \emph{t+1}. Based on this value ($\rho_c$[1] - $\rho_c$[3]), a value of 1.0 is assigned to $\alpha_s[c]$ for the top K\%  of companies (which have a bigger value) and -1.0 is assigned to $\alpha_s[c]$ for the bottom K\% of companies (which have a smaller value). As done in the market neutral portfolio above, we divide each element of $\alpha_s$ by the sum of the absolute value of the element of $\alpha_s$ ($\Sigma_s$ = $\sum_{c=1}^N |\alpha_s[c]| $) and use this as the top/bottom K portfolio. The subscript \emph{s} is added to represent the top/bottom K portfolio. Steps to create the top/bottom K portfolio are described in Algorithm 3.

The main aim of testing the performance of the top/bottom K portfolio is as follows. We used the Q-learning algorithm for training, which uses an action value that corresponds to the expected cumulative reward of an action. In this sense, a larger action value of the long (short) position should indicate more profit the model will receive if the model takes a long (short) position. So if our CNN is trained properly, the top/bottom K portfolio that takes a position based on the subtracted value ($\rho_c$[1] - $\rho_c$[3]) should yield more profit than the market neutral portfolio that distributes asset to all companies. Thus, by this test, we are able to show that our CNN is not only capable of choosing the best action among Long, Neutral, Short actions, but also can assign higher values to more profitable actions. In other words, our CNN can distinguish strong patterns from weak ones.

Table \ref{tab:Table4} compares the overall performance of the top/bottom K portfolio with the average annual market return. The result generally shows that when the portfolio distributes more of its asset to a smaller number of companies that have larger action value, the annual return increases. Although the transaction costs are not included in the result, the result shows that the annual return is much higher than the average market return in most countries for the majority of the time period. The average tendency is also shown in Fig. \ref{fig:Fig2}. The result clearly shows that decreasing K increases the annual return. The result shows that action values can be used to decide what position to take, and larger action values indicate more profit.

\begin{algorithm}
\caption{ market neutral portfolio}
\begin{algorithmic}[1]
\STATE Initialize $\alpha_n$ $\leftarrow$  0 \\

{\setstretch{1.2} 
\FORALL{  \emph{c} }
    \IF{  $\eta_c[1]$ == 1}
        \STATE $\alpha_n[c]$ $\leftarrow$ 1 \\
    \ELSIF{ $\eta_c[3]$ == 1 }
        \STATE $\alpha_n[c]$ $\leftarrow$ -1 \\
    \ENDIF
\ENDFOR
}
{\setstretch{1.3} 
\STATE$\mu_n$ $\leftarrow$ $\frac{1}{N} \sum_{c=1}^N \alpha_n[c]$ \\
\STATE$\alpha_n[c]$ $\leftarrow$ $\alpha_n[c]$ - $\mu_n$  for all \emph{c} \\
\STATE$\Sigma_n$ $\leftarrow$ $ \sum_{c=1}^N |\alpha_n[c]|$ \\
\STATE$\alpha_n[c]$ $\leftarrow$ $\alpha_n[c]$ / $\Sigma_n$ for all \emph{c} \\ 
}
\end{algorithmic}
\end{algorithm}

\begin{table*}[ht]
\caption{Statistical results of the random portfolios compared to our market neutral, top/bottom K portfolios.  }
\label{tab:Table5} 
\centering
\begin{tabular}{|l||l|l|l|l|l|l|l|l|l|l|l|l|}\hline

& \multicolumn{4}{ c |} {\textbf{ N=3000}} 
&\multicolumn{4}{ c |} {\textbf{ N=500}} 
&\multicolumn{4}{ c |} {\textbf{ N=100 }} \\

\hline  Portfolio &   $\mu$  & $\sigma$  &  $\overline \mu$ & Z-score & $\mu$  & $\sigma$ &  $\overline \mu$ &  Z-score & $\mu$  & $\sigma$  &  $\overline \mu$ & Z-score   \\

\hline neutral &  $\approx$ 0  & 0.299 & 11.38 & 38.06 & $\approx$ 0 & 0.982 & 27.16 & 27.66  & $\approx$ 0 & 1.160 & 46.56 & 40.14 \\
\hline K = 20  &  $\approx$ 0  & 0.405 & 18.85 & 46.54 & $\approx$ 0 & 1.355 & 51.86 & 38.27  & $\approx$ 0 & 1.596 & 77.6 &  48.62 \\
\hline K = 10  &  $\approx$ 0  & 0.574 & 28.23 & 49.18 & $\approx$ 0 & 1.904 & 78.64 & 41.30  & $\approx$ 0 & 2.210 & 122.57 & 55.46 \\
\hline K = 5   &  $\approx$ 0  & 0.811 & 40.04 & 49.37 & $\approx$ 0 & 2.696 & 112.1& 41.58  & $\approx$ 0 & 3.149 &  166.95 & 53.02 \\
\hline
\end{tabular}
\end{table*}

\subsection{Statistical tests}

All the portfolios in our experiments are market neutral and essentially have Beta of zero, so we constructed random market neutral portfolio and random top/bottom K portfolios. We compare their results with those of our portfolios to verify the statistical significance of our results. Since the number of companies and the type of portfolio affect the standard deviation of the portfolio return, each of the 4 portfolio types (neutral, K=20, 10, and 5) was tested in 3 types of countries (initial N=3000, 500, and 100) over entire testing period. 10,000 simulations were conducted for each experiment, and the mean $\mu$ and standard deviation $\sigma$ of the annual return were calculated. Random portfolios are created as follows. The value [-1,1] was randomly selected, neutralized, and divided by the sum of absolute values to construct the random market neutral portfolio $\alpha_n$ using the same method mentioned in the market neutral portfolio section. So $\alpha_n$ is a randomly weighted portfolio which satisfies $ \sum_{c=1}^N \alpha_n[c] = 0 $ and $ \sum_{c=1}^N |\alpha_n[c]| = 1.0 $. The random top/bottom K portfolio $\alpha_s$ was also generated in a similar way. K\% of randomly selected companies took a long position and the other K\% of randomly selected companies took a short position. Like the portfolio of our model, both random portfolios were also reconstructed every day.

A portfolio return is assumed to be normally distributed, so we calculated Z-scores for the statistical test. The results are provided in Table \ref{tab:Table5}. In Table \ref{tab:Table5}, $\mu$ and $\sigma$ indicate the mean and the standard deviation of the annual return of the random portfolios, respectively. Since all the four random portfolios (neutral, K=20, 10, 5) are actually market neutral, $\mu$ is zero. The standard deviation $\sigma$ tends to increase as K and N become smaller; K and N become smaller when the portfolio distributes its asset to a smaller number of companies. Also in Table \ref{tab:Table5}, $\overline \mu$ indicates the average annual return of our corresponding portfolio in a given type of country over entire testing period (12 years). For example, in the experimental result of the neutral portfolio with an N value of 3000 provided in Table \ref{tab:Table5}, $\overline \mu$ shows that the market neutral portfolio in the US yields a 11.38 $\approx(20.64+9.39+4.11)/3$ percent return per annum over 12 years. Thus, we obtain a Z-score ( $(\overline \mu - \mu)/\sigma$) of 38.06 $\approx (11.38-0)/0.299$ for the market neutral portfolio tested in US. In this way, we are able to compare our portfolios with the random portfolios and obtain statistical results. As the result shows, in most cases, mean of our portfolio returns are usually more than 30-$\sigma$ away from the mean of the random portfolio returns.

\begin{algorithm}
\caption{ top/bottom K portfolio}
\begin{algorithmic}[1]
\STATE Initialize $\alpha_s$ $\leftarrow$  0 \\

{\setstretch{1.2} 
\FORALL{ \emph{c} }
    \IF{   $\rho_c[1]-\rho_c[3]$ is in the top K\% }
        \STATE $\alpha_s[c]$ $\leftarrow$ 1 \\
    \ELSIF{ $\rho_c[1]-\rho_c[3]$ is in the bottom K\%  }
        \STATE $\alpha_s[c]$ $\leftarrow$ -1 \\
    \ENDIF
\ENDFOR
}
{\setstretch{1.3} 
\STATE $\Sigma_s$ $\leftarrow$ $ \sum_{c=1}^N |\alpha_s[c]|$ \\
\STATE $\alpha_s[c]$ $\leftarrow$ $\alpha_s[c]$ / $\Sigma_s$ for all \emph{c} \\ 
}
\end{algorithmic}
\end{algorithm}

\section{Conclusion}
Regardless of the problem domain, the generality of a model is always an important aspect to consider when selecting a model for a problem. For example, can a QA system that is trained to answer questions on natural science articles also correctly answer social science questions? Can a face recognition system trained on images of European faces also classify images of Asian faces? These kinds of questions are not only important for industrial solutions, but they are also very interesting research topics. We conducted numerous experiments to determine whether our model trained on certain patterns in stock charts from a single country can make profit not only in the given country but also in other countries. As our results show, our model trained in only the US market, also performed well or even better in many other markets for the 12-year testing period. 

Although the result shows extraordinary annual returns in some countries, we are not insisting that implementing our model would achieve the exact same amount of profit as shown. In real practice, transaction fees or taxes, and the bid-ask spread or actual volume of stock the investor could trade at a certain price should be considered. Yet, our result strongly indicates that people, regardless of culture or country, react similarly to certain past price/volume patterns. Based on this observation, artificial intelligence and machine learning based stock price forecasting studies, which have been conducted in only a single country so far, can be employed in global stock market. In other words, if the model structure, input feature, and training procedure are proper, training and testing do not need to be done in the same market. 

\section*{Acknowledgment}

This work was supported by the National Research Foundation of Korea   (NRF-2017R1A2A1A17069645, NRF-2017M3C4A7065887).

\end{document}